# Squeezed-Light-Enhanced Dispersive Gyroscope based Optical Microcavities

XIAOYANG CHANG,[1] WENXIU LI[1], HAO ZHANG[2*], YANG ZHOU[2], ANPING HUANG[1], AND ZHISONG XIAO[1,2,3]

*1 School of Physics, Beihang University, Beijing 100191, China*
*2 Research Institute of Frontier Science, Beihang University, Beijing 100191, China*
*3 Beijing Academy of Quantum Information Sciences, Beijing 100193, China*
*\*Corresponding author: haozhang@buaa.edu.cn*

**Abstract:** Optical gyroscope based on the Sagnac effect have excellent potential in the application of high-sensitivity inertial rotation sensors. In this paper, we demonstrate that for an optical resonance gyroscope with normal dispersion, the measurement sensitivity can be increased by two orders of magnitude through coupling into a squeezed vacuum light, which is different from that in the classical situation. When the system is operated under critical anomalous dispersion condition, injecting a squeezed vacuum light allows the measurement sensitivity beyond the corresponding standard quantum limit by five orders of magnitude, with a minimum value of $3.8 \times 10^{-5}$ Hz. This work offers a promising possibility for developing optical gyroscopes that combine high sensitivity with tiny size.

## 1. Introduction

Optical gyroscopes based on microcavity have remarkable advantages in weight, size, cost, and power consumption and have become one of the focuses of inertial rotation sensors[1]. Physically, the optical gyroscope is based on the Sagnac effect[2]. That is to say, if two beams are counter-propagating in an optical loop, a phase or frequency shift between the clockwise (CW) and counterclockwise (CCW) beams is generated when the loop is rotating around the axis, which is proportional to the angular rotation rate of the loop. The accumulated phase or frequency shift in the loop, which comes from the Sagnac effect, is proportional to the effective area of the optical cavity. Therefore, the Sagnac phase shift accumulated in an optical microcavity gyro is seriously bounded, limiting its measurement sensitivity due to the tiny effective area. In many applications such as vehicles, spacecraft, and satellites, the size and weight of the gyro are strictly limited, but high sensitivity is required simultaneously.

In order to make a microcavity optical gyro have high measurement sensitivity, various coupled resonator optical waveguides (CROWs) were proposed, and the normal dispersion effect in these structures is also discussed[3–8]. Subsequent studies have shown that even with optimized parameters, the sensitivity of the CROW gyro is equal to that of a Ring laser gyro under equal loop loss conditions. Physically, normal dispersion is a resultant feature of the CROW structure and is not directly correlated with sensitivity enhancement. In addition, the sensitivity of a resonator-based optical gyroscope can be enhanced by introducing anomalous dispersion effect[9]. Typically, there are two experimental regimes for achieving anomalous dispersion, utilizing the alkali metal vapors and optical coupled resonators[9–24]. However, the alkali metal vapors approach is unsuitable for reducing the weight and size of a gyro because the experimental system is rather complex. Optical coupled microresonators can achieve anomalous dispersion only by using passive elements, thus avoiding the complexities of modulating the dispersion in alkali metal vapors.

Nevertheless, most of the previous research on dispersion enhancing the sensitivity of an optical gyroscope is based on classical light. As a result, these devices are forcing an inevitable quantum-mechanical limit on the sensitivity in detecting angular rotation rate[25]. David D. Smith et al. have shown that a cavity containing rubidium atomic gas dispersion medium, the sensitivity to resonant frequency shift can be increased by two orders of magnitude at critical anomalous dispersion point under ideal conditions (classical sources in the cavity can be neglected, e.g., temperature and mechanical fluctuations) [26]. This increase above corresponds to the standard quantum limit (SQL) dispersion enhancement of the system due to the bound of the quantum-mechanical photon shot noise. Fortunately, the SQL is a limit that can be reached when the measurement device uses classical light, but it is not the limit that theoretical is allowed. Besides, the SQL can be exceeded by using non-classical light[27], such as the Fock states[28], the NOON states[29], and the squeezed states of light. In particular, squeezed vacuum light has a particular advantage in exceeding the SQL due to its intrinsic nature of the noise distribution on the two quadrature components[30–32].

In this paper, we present a scheme that a squeezed vacuum light coupling into a three-dimensional vertically coupled resonator system (3D-VCRS) with dispersion to improve the measurement sensitivity on angular rotation rate. The rest of this article is organized as follows. In Section 2, we calculated the dispersion condition and derived the expression of scale factor enhancement and the uncertainty of frequency measurement of the system. In Section 3, we discussed the sensitivity enhancement factor and the measurement sensitivity of the system operating under different situations. In Section 4, we conclude with a summary.

## 2. Theoretical mode and analysis

A schematic view of squeezed vacuum light coupling with the 3D-VCRS is shown in Fig. 1. The system consists of two ring resonators, Ring 1 and 2, and an input waveguide. A coherent light $\hat{a}_c$ passes through a 50/50 beam splitter (BS) and launches into the input waveguide at the bottom of the system, then couples to the stacked vertically coupled ring resonators. A squeezed vacuum light (quadrature phase squeezed light) $\hat{b}_{in}$ is reflected by the BS and coupled into the 3D-VCRS together with the coherent light $\hat{a}_c$ as input light $\hat{a}_{in}$. A blocker is placed between the squeezed light source and the BS, and the blocker is opened or closed, corresponding to $\hat{b}_{in}$ is squeezed vacuum light or vacuum field. The output light $\hat{a}_{out}$ is detected by a photodetector (PD). The advantage of 3D-VCRS is that the input light $\hat{a}_{in}$ is transmitted in the same direction in both Ring 1 and 2, which allows the Sagnac phase shift to accumulate. Thus avoid the cancellation of the Sagnac phase shift in the unfolded LC coupled ring resonators structure[33]. We assume that the BS and the input waveguide are lossless and that the loss coefficients of Ring 1 and 2 are $a_1$ and $a_2$, respectively. $r_1$, $r_2$, and $t_1$, $t_2$ are the amplitude reflection and transmission coefficients in the two coupling regions, i.e., the coupling region between Ring 1 and 2, and the region between Ring 2 and the input waveguide, respectively. Here, we neglect the loss in the coupling region, and it can be shown that $r_j^2 + t_j^2 = 1$ (i=1, 2).

The complex transmission coefficient in the coupling region between Ring 2 and 1 can be written as[34]

$$\frac{E_2}{E_0} = \frac{r_1 - a_1 \exp(i\phi_1)}{1 - a_1 r_1 \exp(i\phi_1)} = \tilde{t}_1(\omega) = |t_1(\omega)| e^{i\Phi_1}, \tag{1}$$

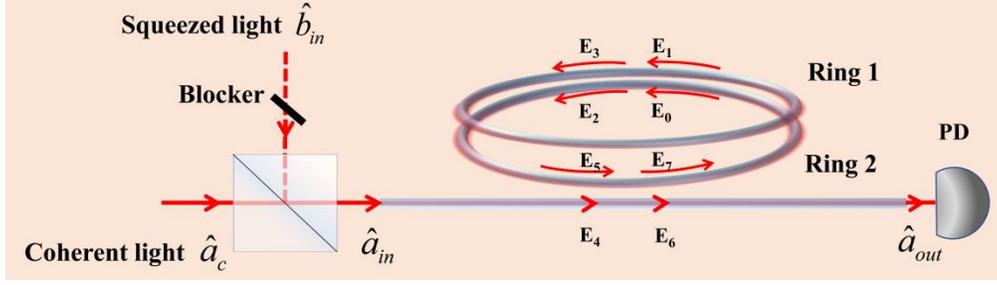

Fig. 1. Schematic of squeezed light coupling with the 3D-VCRS structure

here $\phi_1 = \dfrac{2\pi r_{R1} \cdot \omega \cdot n_{eff,R1}}{c}$ is the one round trip phase shift in Ring 1, where $r_{R1}$, $n_{eff,R1}$, $\omega$, and $c$ are the radius, the effective index of Ring 1, angular frequency of the cavity mode, and the speed of light in vacuum, respectively. The complex transmission coefficient of the 3D-VCRS is

$$\frac{E_6}{E_4} = \frac{r_2 - a_2 \tilde{t}_1(\omega)\exp(i\phi_2)}{1 - a_2 r_2 \tilde{t}_1(\omega)\exp(i\phi_2)} = \tilde{\tau}(\omega) = |\tau(\omega)|e^{i\Phi}, \quad (2)$$

here $\phi_2 = \tau_c\left(\omega_q - \omega_q^{(e)}\right)$ is the one round trip phase shift in Ring 2, where $\omega_q$ and $\omega_q^{(e)}$ are the qth cavity mode frequency of Ring 2 with and without dispersion, respectively[35]. $\tau_c$ is the round trip time of light in Ring 2. When Ring 1 is removed (empty cavity without dispersion), the transmission coefficient in the coupling region between Ring 2 and the input waveguide can be written as

$$\tilde{t}_2(\omega) = \frac{r_2 - a_2 \exp(i\phi_2)}{1 - a_2 r_2 \exp(i\phi_2)} = |t_2(\omega)|e^{i\Phi_2}. \quad (3)$$

The effective group index $n_g$ of the 3D-VCRS can be given as

$$n_g = n_{eff,R2} + \frac{c}{L_{R2}}\frac{\partial \Phi_1}{\partial \omega}, \quad (4)$$

here the effective length $L_{R2}$ is $2\pi r_{R2}$ ($r_{R2}$ is the radius of Ring 2), and the effective index of Ring 2 is $n_{eff,R2}$, $\Phi_1$ is the argument of $\tilde{t}_1(\omega)$.

For a passive coupled resonator, it exhibits normal dispersion ($1<n_g$) when it is over-coupled and anomalous dispersion ($0<n_g<1$) when it is under-coupled, which can be achieved by tuning the coupled coefficient between Ring 1 and 2, i.e., by tailoring the value of $r_1$. In order to determine the detailed parameters with which the system operates under normal or anomalous dispersion conditions, $n_g$ calculated from Eq. (4) versus $\Delta\omega/2\pi$ with different values of $r_1$ is plotted in Fig. 2. $\Delta\omega$ is the detuning between the angular frequency of input light and the resonance frequency of the 3D-VCRS. The fixed values are $L_{R1} = L_{R2} = 50\lambda_0/n_{eff,R}$, $\lambda_0 = 1064nm$, $n_{eff,R1}=n_{eff,R2}=2$, $a_1 = 0.94$, $a_2 = \exp(-2\pi r_{R2}\alpha_2)$, where $\alpha_2 = 3\ dB/cm$ is the propagation loss in Ring 2.

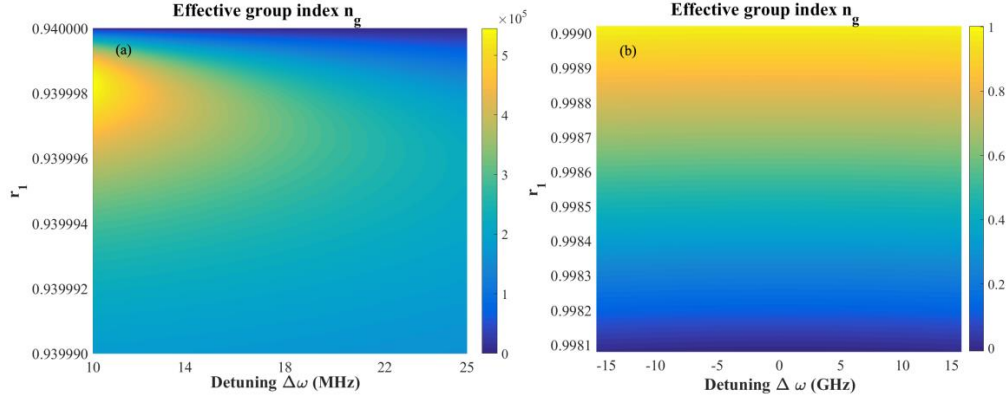

Fig. 2. The effective group index $n_g$ when $L_{R1} = L_{R2} = 50\lambda_0 / n_{eff,R}$, $\lambda_0 = 1064 nm$, $n_{eff,R1}=n_{eff,R2}=2$, $a_1 = 0.94$, $a_2 = \exp(-2\pi r_{R2}\alpha_2)$, $\alpha_2 = 3\ dB/cm$. **(a)** normal dispersion; **(b)** anomalous dispersion.

As shown in Fig. 2 (a), if $r_1$ is between 0.9399 and 0.94, the value of $n_g$ is much greater than one, meaning the system operates under normal dispersion conditions. Furthermore, the pole of $n_g$ can be obtained at $r_1$=0.93999. In Fig. 2 (b), we show the situation of anomalous dispersion. When $r_1$ changes from 0.9981 to 0.999, the value of $n_g$ satisfies $0<n_g<1$. It should be noted that if $r_1$ =0.99809, the value of $n_g$ approaches zero at near resonance, which is the critically anomalous dispersion condition of this 3D-VCRS.

For a passive optical cavity, the frequency shift caused by an external disturbance will change if a dispersion element is introduced; if the dispersion is normal (or anomalous), the frequency shift will be reduced (or increased). The scale factor is defined as the frequency shift caused by an external disturbance (e.g., rotation or change in cavity length). The scale factor enhancement of the system is given by

$$S_{en}(\omega) = \frac{d\omega_q}{d\omega_q^{(e)}} = \left[1 + \frac{1}{\tau_c}\frac{d\Phi_1}{d\omega} + \frac{1}{\tau_c}\frac{dF}{d\omega}\right]^{-1}$$
$$= \left[\frac{n_g(\omega)}{n_{eff,R}} + T_{cav}(\omega)\right]^{-1}, \quad (5)$$

here $F$ is the additional phase factor that comes from the mode reshaping, and can be written as

$$F(\omega) = \arcsin\frac{E(\omega)}{\sqrt{E(\omega)^2 + G(\omega)^2}} - \arcsin\frac{D(\omega)}{\sqrt{E(\omega)^2 + G(\omega)^2}}. \quad (6)$$

Where

$$D = \left[\begin{array}{l}(g-1)\left[2g'\left(a_2^2 t_1^2 + t_2^2 - g\right) + 2a_2^2 t_1 t_1'(1-g)\right] \\ +4g\left(gg' - a_2^2 t_1 t_1'\right) - 2g'\left(g^2 - t_2^2 - a_2^2 t_1\right)\end{array}\right],$$
$$E = 2\left[g' t_2^2 + g' a_2^2 t_1^2 - 2a_2^2 t_1 t_1' g\right],$$
$$G = 2g\left(g^2 - t_2^2 - a_2^2 t_1^2\right)n_g,$$
$$g = r_2 a_2 |t_1|. \quad (7)$$

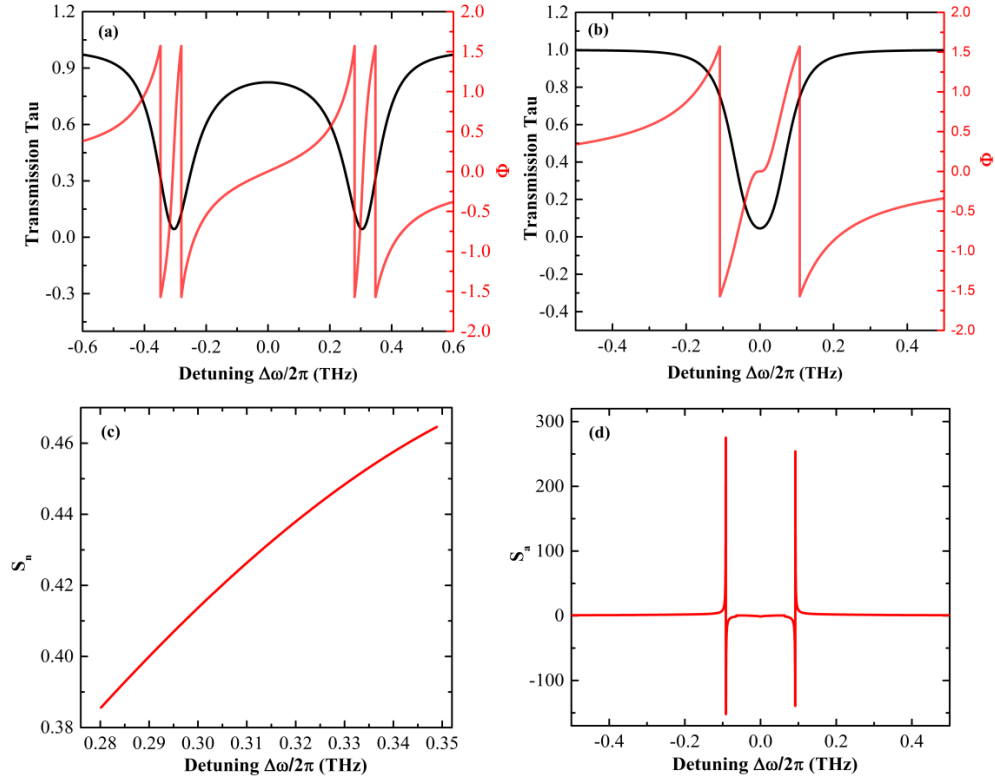

Fig. 3. The transmission characteristics (a) and (b), as well as the scale factor enhancement (c) and (d) of the 3D-VCRS under different coupled coefficient conditions. In Fig. 3 (a) and (c), $r_1$=0.93999; In Fig. 3 (b) and (d), $r_1$=0.99809, the other parameters are identical to their values in Fig. 2

$|t_1|$ is the modulus of $\tilde{t}_1$, $g' = dg(\omega)/d\omega$. $T_{cav}(\omega) = \frac{1}{\tau_c}\frac{dF}{d\omega}$ is the corresponding additional dimensionless time delay.

Now, let us consider the transmission characteristics of the 3D-VCRS under different coupled coefficient conditions. Fig. 3 (a) shows the cavity transmittance (black) and the total phase shift (red) calculated from Eq. (2) as a function of the frequency detuning. The coupled coefficient is set to $r_1$=0.93999, and the other parameters are identical to that in Fig. 2. We can see two dips distributed at the positive and negative frequency detuning of the resonant frequency. Each dip corresponds to a frequency region where the total phase shift varies sharply with frequency detuning. The derivative of the total phase shift with respect to frequency detuning is positive, indicating that the system is under normal dispersion conditions. In Fig. 3 (b), we show the case when $r_1$=0.99809. As we can see that the transmission has an apparent dip at zero detuning. Furthermore, two frequency regions are distributed symmetrically at the near-resonant frequency, where the total phase shift varies dramatically to the frequency detuning. The derivative of the total phase shift to the frequency detuning is negative at the two detuning regions above, showing that the system is operated in anomalous dispersion conditions.

In Fig. 3 (c), $r_1$=0.93999, we plot the scale factor enhancement $S$ in the normal dispersion frequency region at the positive detuning in Fig. 3 (a). It can be seen that the value of $S$ is less than one during this frequency region, meaning that the resonant frequency shift of the system becomes smaller than that in the situation without dispersion. This result suggests that normal dispersion reduces the measurement sensitivity of the resonant frequency shift of an optical cavity. In Fig. 3 (d), $r_1$=0.99809, $S$ at the identical frequency region in Fig. 3 (b) is plotted. We can see two poles of S occur at the corresponding anomalous dispersion frequency region. The pole of S suggests that the resonant frequency shift can be increased by introducing anomalous dispersion compared to the situation without dispersion.

For an optical resonator without dispersion, the frequency difference in the two count-propagation lights can be written as

$$\Delta f = \frac{4A}{L_R \lambda} \Omega, \qquad (8)$$

here $A$, $L_R$, $\Omega$, and $\lambda$ are the enclosed area, the effective length, the rotational angular velocity of the resonator, and the wavelength of the input light, respectively. If the resonator is operating under dispersion conditions, the frequency difference can be rewritten as

$$\Delta f = S_{en} \frac{4A}{L_R \lambda} \Omega. \qquad (9)$$

$S_{en}$ is the scale factor enhancement induced by dispersion. We define the minimum measurable angular velocity as the measurement sensitivity for an optical gyro. According to Eq. (9), the measurement sensitivity of a gyro with dispersion is given by

$$\delta\Omega_{min} = \Delta f_{min} \frac{1}{S_{en}} \frac{L_R \lambda_0}{4A}, \qquad (10)$$

$\Delta f_{min}$ is the minimum measurable frequency. From Eq. (10), for a given optical resonator with dispersion, since the enclosed area, the effective length, and the wavelength of the input light are fixed, the measurement sensitivity can be obtained by simply calculating the scale factor enhancement and the minimum measurable frequency. The expression of scale factor enhancement has been derived above, and we now calculate the minimum measurable frequency of the system.

The photon number and phase of a light obey the uncertainty relation and can be expressed as[36]

$$\Delta n \Delta \Phi \geq \frac{1}{2}, \qquad (11)$$

here $\Delta n$ is the standard deviation of the photon number, and $\Delta \Phi$ is the standard deviation of the phase. This relationship yields

$$\Delta \Phi = \frac{1}{2\Delta n}. \qquad (12)$$

The resonance frequency of a cavity mode cannot be determined to exceed the accuracy permitted by the phase uncertainty. We assume that the photon lifetime in an optical resonator is $t$, the uncertainty in frequency can be written as

$$\Delta \omega_t = \frac{\Delta \Phi_t}{t} = \frac{1}{\sqrt{2t^2 (\Delta n)^2}}. \qquad (13)$$

Eq. (13) can be expressed in terms of the measurement bandwidth $B$

$$\Delta \omega_B = \sqrt{\frac{B}{t(\Delta n)^2}}. \qquad (14)$$

From Eq. (13), when the photon lifetime $t$ in the resonator and the measurement bandwidth $B$ of a PD are determined, the frequency measurement uncertainty depends on the photon number variance of the output light.

Next, let us analyze the photon number variance of the output light for the 3D-VCRS. The quantum states we consider here are described with annihilation and creation operators ($\hat{a}$ and $\hat{a}^\dagger$) of the electromagnetic field. The canonical commutation relation is $\left[\hat{a}, \hat{a}^\dagger\right] = 1$. We implement the Fourier transformation; the annihilation and creation operators can be expressed as $\hat{a}(\Omega) = \frac{1}{\sqrt{2\pi}} \int dt \hat{a}(t) e^{-i\Omega t}$, $\hat{a}^\dagger(\Omega) = \frac{1}{\sqrt{2\pi}} \int dt \hat{a}^\dagger(t) e^{-i\Omega t}$ in the frequency domain. Considering a 50/50 BS, the input light $\hat{a}_{in}$ coupling into the input waveguide can be written as

$$\hat{a}_{in} = \frac{1}{\sqrt{2}} \left( \hat{a}_c + i\hat{b}_{in} \right),$$
$$\hat{a}_{in}^\dagger = \frac{1}{\sqrt{2}} \left( \hat{a}_c^\dagger - i\hat{b}_{in}^\dagger \right). \tag{15}$$

According to Eq. (2), the complex transmission coefficient of the 3D-VCRS is $\tilde{\tau}$, then the output light of the system is given by

$$\hat{a}_{out} = \hat{a}_{in} \tilde{\tau}$$
$$\hat{a}_{out}^\dagger = \hat{a}_{in}^\dagger \tilde{\tau}^*. \tag{16}$$

Therefore, we can obtain the mean number of photons of the output light as

$$a_{out}^\dagger a_{out} = a_{in}^\dagger a_{in} |\tau|^2 = \frac{1}{2} |\tau|^2 \left( a_c^\dagger a_c + i a_c^\dagger b_{in} - i b_{in}^\dagger a_c + b_{in}^\dagger b_{in} \right). \tag{17}$$

Here the first term is the mean photon number of the coherent light $\hat{a}_c$, the second and third terms are due to the interference between mode $\hat{a}_c$ and $\hat{b}_{in}$, and the fourth term is the mean photon number of mode $\hat{b}_{in}$. Coherent state is the eigenstate of the annihilation operator $\hat{a}_c$, the eigenvalue $a_c = |a_c| e^{i\theta}$ is a complex number, and the mean number of photons is $\langle a_c^\dagger a_c \rangle = |a_c|^2$. For simplicity, we assume $\theta = 0$. According to Eq. (17), the photon number variance of the output light can be expressed as

$$(\Delta n)^2 = \left\langle \left( a_{out}^\dagger a_{out} \right)^2 \right\rangle - \left\langle \left( a_{out}^\dagger a_{out} \right) \right\rangle^2$$
$$= \frac{1}{4} |\tau|^4 |a_c|^2 \left( b_{in} b_{in}^\dagger - b_{in} b_{in} + b_{in}^\dagger b_{in} - b_{in}^\dagger b_{in}^\dagger \right). \tag{18}$$

When mode $\hat{b}_{in}$ is a vacuum field, the mean photon number is zero, i.e., $\langle b_{in}^\dagger b_{in} \rangle = 0$, and the second and fourth terms in Eq. (18) are zero. In this situation, Eq. (18) can be written as

$$\left( \Delta n_{vaccum} \right)^2 = \frac{1}{4} |\tau|^4 |a_c|^2. \tag{19}$$

The subscript "vacuum" represents the case that mode $\hat{b}_{in}$ is a vacuum field. If mode $\hat{b}_{in}$ is a squeezed vacuum light, which can be described as a quantum field with the uncertainty on one quadrature phase less than vacuum fluctuation and the fluctuation on the other quadrature phase greater than vacuum noise. With squeezed light, the measurement sensitivity can be improved significantly by encoding the signal to be detected in the noise-reduced quadrature phase. The mean photon number is $\langle b_{in}^\dagger b_{in} \rangle = \sinh^2 s$; here, $s$ is the squeeze factor, and in this paper, we assume $s = 5$[37]. The photon number variance of the output light can be written as

$$\left( \Delta n_{squeezed} \right) = \frac{1}{4} |\tau|^4 |a_c|^2 \left( \sinh 2s + \cosh 2s \right). \tag{20}$$

The subscript "squeezed" denotes that a squeezed vacuum light is injected into the system. We introduce the mean power $P = \frac{\hbar \omega \langle n \rangle}{t}$; here, $\langle n \rangle$ represents the mean number of photons of the input light. The quality factor Q of an optical resonator is related to the mean power decay time $t$ by $\frac{1}{t} = \frac{\omega_0}{Q}$. Using Eqs. (19), (20), and (14) we can evaluate the uncertainty of frequency measurement for the system:

$$\Delta\omega_B = \sqrt{\frac{B\hbar\omega}{P\mu}}\frac{\omega_0}{Q}. \tag{21}$$

$\mu$ can be classified into three different situations:

$$\mu_{empty} = \frac{1}{4}|t_2|^4,$$

$$\mu_{d,vacuum} = \frac{1}{4}|\tau|^4, \tag{22}$$

$$\mu_{d,squeezed} = \frac{1}{4}|\tau|^4 (\sinh 2s + \cosh 2s).$$

The subscript "empty" indicates that the system operates in an empty cavity with no dispersion, and the squeezed light is blocked. The subscript "d, vacuum" denotes the system with dispersion but without a squeezed vacuum light injecting. The subscript "d, squeezed" represents the situation where the system operates in a dispersion condition, and a squeezed vacuum light is injected. Using Eqs. (5), (10), (21), and (22), we obtain

$$\delta\Omega_{empty} = \sqrt{\frac{B\hbar\omega}{P\mu_{empty}}}\frac{\alpha_{eff}c\lambda_0}{2L}, \tag{23}$$

$$\delta\Omega_{d,vacuum} = \frac{1}{S_{en}}\sqrt{\frac{B\hbar\omega}{P\mu_{d,vacuum}}}\frac{\alpha_{eff}c}{n_g}\frac{\lambda_0}{4L}, \tag{24}$$

$$\delta\Omega_{d,squeezed} = \frac{1}{S_{en}}\sqrt{\frac{B\hbar\omega}{P\mu_{d,squeezed}}}\frac{\alpha_{eff}c}{n_g}\frac{\lambda_0}{4L}, \tag{25}$$

Eqs. (23), (24), and (25) correspond to the minimum measurable rotational angular velocity of the 3D-VCRS, i.e., the measurement sensitivity, in three different situations, respectively.

### 3. Sensitivity enhancement factor and measurement sensitivity

In this paper, we define the ratio of the measurement sensitivity of the system when it is with dispersion to that without dispersion as the sensitivity enhancement factor. Using Eqs. (23) and (24), the sensitivity enhancement factor when squeezed vacuum light is blocked can be written as

$$\chi_{d,vacuum} = \frac{\delta\Omega_{d,vacuum}}{\delta\Omega_e} = \frac{1}{2}\frac{1}{S_{en}n_g}\sqrt{\frac{\mu_{empty}}{\mu_{d,vacuum}}}, \tag{26}$$

If a squeezed vacuum light is injected into the system, using Eqs. (23) and (25), the sensitivity enhancement factor can be written as

$$\chi_{d,squeezed} = \frac{\delta\Omega_{d,squeezed}}{\delta\Omega_e} = \frac{1}{2}\frac{1}{S_{en}n_g}\sqrt{\frac{\mu_{empty}}{\mu_{d,squeezed}}}. \tag{27}$$

First, let us consider the situation when squeezed vacuum light is blocked. Fig. 4 (a) shows the sensitivity enhancement factor $\chi_{n,vacuum}$ (subscript "n" indicates normal dispersion), computed from Eq. (26), when the system operates in the normal dispersion frequency region. It can be seen that there is a peak of $\chi_{n,vacuum}$ at the positive frequency detuning around 0.305 THz, with a peak value of about twelve. The result shows that the measurement sensitivity of the system is reduced by introducing normal dispersion, which is consistent with previous studies. If the system is operating in the anomalous dispersion frequency region, the sensitivity enhancement factor $\chi_{a,vacuum}$ (subscript "a" indicates anomalous dispersion) is plotted in Fig. 4 (b). We can see that the minimum value of $\chi_{\min,a,vacuum} = 0.0019$ occurs at the positive detuning of about 0.09 THz. This pole of $\chi_{\min,a,vacuum}$ is because, at the frequency detuning around 0.09 THz, the system approaches the critical anomalous dispersion condition, where the associated powerful mode pushing effect increases the resonance frequency shift dramatically, leading to an increase in the measurement sensitivity.

Secondly, we consider the case when a squeezed vacuum light is coupled into the system. In Fig. 4 (c), we plot the sensitivity enhancement factor $\chi_{n,squeezed}$ when the system is under normal dispersion conditions. Similar to the situation in Fig. 4 (a), $\chi_{n,squeezed}$ also occurs a peak at the positive detuning around 0.305 THz. However, the maximum value is $\chi_{n,squeezed} = 0.0804$, much smaller than that in Fig. 4 (a). This decrease in $\chi_{n,squeezed}$ is because, although the mode pulling effect associated with normal dispersion reduces the resonant frequency shift and thus introduces negative feedback. When a squeezed vacuum light is coupled into the system, the noise in the quadrature phase is squeezed, causing a reduction in the uncertainty of frequency measurement of the cavity mode, thereby generating positive feedback on the measurement sensitivity. The positive feedback introduced by squeezed vacuum light is larger than the negative feedback associated with normal dispersion, thus producing a noticeable net enhancement in the measurement sensitivity of the system.

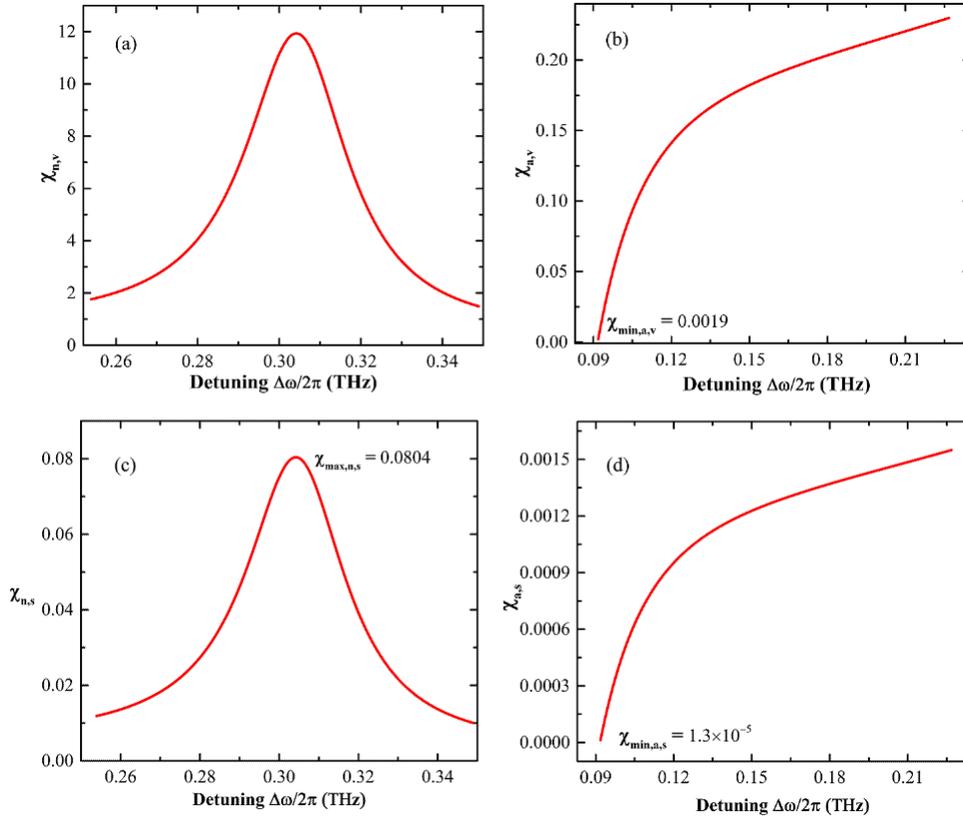

Fig. 4. The sensitivity enhancement factor: (a) and (b) corresponding to the system operating at normal and anomalous dispersion conditions, respectively when squeezed vacuum light is blocked; (c) and (d) corresponding to the case for normal and anomalous dispersion when a squeezed vacuum light is injecting.

Fig. 4 (d) shows the sensitivity enhancement factor $\chi_{a,squeezed}$ when the system is under anomalous dispersion conditions. We can see that the minimum value of $\chi_{\min,a,squeezed} = 1.3 \times 10^{-5}$ occurs near the frequency detuning of 0.09 THz. This dramatic decrease in $\chi_{\min,a,squeezed} = 1.3 \times 10^{-5}$ is because when anomalous dispersion is introduced into the system, the associated mode pushing effect greatly increases the resonant frequency shift, which provides positive feedback on the measurement sensitivity. In addition, when a squeezed vacuum light is injected into the system, another positive feedback, as described above, will generate. The positive feedback from anomalous dispersion and the squeezed vacuum light are superimposed at the critical anomalous dispersion condition, thus dramatically increasing the measurement sensitivity of the 3D-VCRS. This result demonstrates that when the system operates in anomalous dispersion conditions, the measurement sensitivity can be dramatically increased by coupling into a squeezed vacuum light, even exceeding the corresponding SQL by five orders of magnitude under critical anomalous dispersion condition.

Next, let us consider the measurement sensitivity of the 3D-VCRS under different situations: (i) the system is operating in an empty cavity, and the squeezed vacuum light is blocked; (ii) the system with dispersion but without a squeezed vacuum light injection; (iii) the system operates in dispersion conditions, and a squeezed vacuum light is injected into it simultaneously.

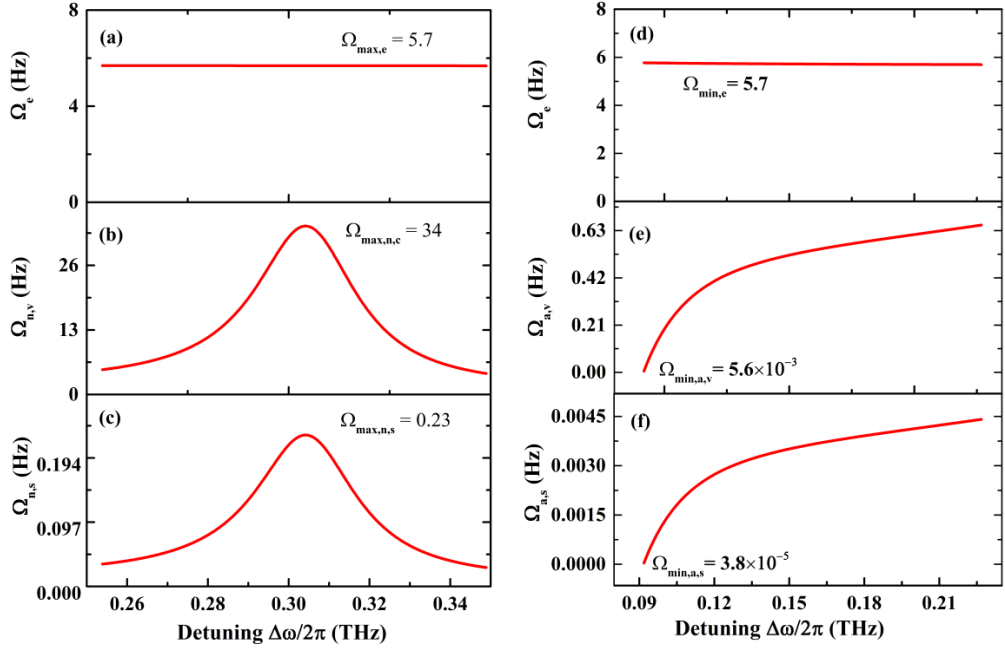

Fig. 4. The measurement sensitivity of the 3D-VCRS under normal dispersion frequency region **(a)** –**(c);** and anomalous dispersion frequency region **(d)** –**(f)**.

Fig. 4 (a), (b), and (c) show the measurement sensitivity of the 3D-VCRS, computed from Eqs. (23), (24), and (25) in the normal dispersion frequency region. In Fig. 4 (a), the measurement sensitivity $\delta\Omega_{empty}$, i.e., in case (i), is plotted. As can be seen that the value of $\delta\Omega_{empty}$ is about 5.7 Hz. Since we have ignored the classical noise in calculations, the result $\delta\Omega_{empty} = 5.7 Hz$ corresponds to the SQL measurement sensitivity of this system. The measurement sensitivity $\delta\Omega_{n,vacuum}$ for case (ii) is plotted in Fig. 4 (b). An apparent peak can be seen near the frequency detuning of 0.305 THz, with a peak value of $\delta\Omega_{\max,n,vacuum} = 34 Hz$, which is larger than $\delta\Omega_{empty} = 5.7 Hz$ in the situation (i). This is because near the 0.305 THz frequency detuning corresponds to a region where the normal dispersion is very strong, and the accompanying negative feedback reduces the measurement sensitivity. Fig. 4 (c) plotted the measurement sensitivity $\delta\Omega_{d,squeezed}$, i.e., the situation (iii). Similar to the case in Fig. 4 (b), $\delta\Omega_{d,squeezed}$ also occurs a peak near 0.305 THz frequency detuning, but the difference is that the peak value is $\delta\Omega_{\max,d,squeezed} = 0.23 Hz$, which is much smaller than $\delta\Omega_{\max,n,vacuum} = 34 Hz$, and smaller than $\delta\Omega_{empty} = 5.7 Hz$. This result indicates that, unlike the classical case (normal dispersion reduces the measurement sensitivity), even though the 3D-VCRS operates under normal dispersion conditions, the measurement sensitivity can be effectively improved by coupling into a squeezed vacuum light.

In Fig. 4 (d), (e), and (f), the measurement sensitivity corresponding to cases (i), (ii), and (iii) are plotted when the 3D-VCRS is operating in the anomalous dispersion frequency region. Fig. 4 (d) shows the measurement sensitivity in case (i). Similarly to the case in Fig. 4 (a), the value of $\delta\Omega_{empty}$ is approximately 5.7 Hz. In Fig. 4 (e), the measurement sensitivity achieves a minimum $\delta\Omega_{\min,a,vacuum} = 5.6 \times 10^{-3} Hz$ at the frequency detuning of around 0.09 THz, which is substantially smaller than $\delta\Omega_{empty} = 5.7 Hz$, The results above demonstrated that the measurement sensitivity of the 3D-VCRS can indeed be enhanced by anomalous dispersion. When the system operates in the anomalous dispersion frequency region and a squeezed vacuum light is injected simultaneously, the measurement sensitivity $\delta\Omega_{a,squeezed}$ is plotted in Fig. 4 (f). Similar to the situation in Fig. 4 (e), the minimum $\delta\Omega_{\min,a,squeezed} = 3.8 \times 10^{-5} Hz$ occurs near the frequency detuning about 0.09 THz, which is significantly smaller than $\delta\Omega_{empty} = 5.7 Hz$ in Fig. 4 (d) and $\delta\Omega_{\min,a,vacuum} = 5.6 \times 10^{-3} Hz$ in Fig. 4 (e). This result demonstrates that a microcavity optical gyro can achieve high sensitivity compared to traditional situations by tailoring the coupled resonator to operate at the critical anomalous dispersion condition and injecting a squeezed vacuum light.

## 4. Conclusion

In this paper, we theoretically investigated the sensitivity enhancement of a microcavity optical gyro with dispersion and a squeezed vacuum light. Our results show that, unlike the classical situation, even though the system operates under normal dispersion conditions, a net increase in measurement sensitivity can be obtained when coupling into a squeezed vacuum light. In addition, with squeezed vacuum light, the system can exceed the corresponding SQL measurement sensitivity by five orders of magnitude at the critical anomalous dispersion condition, obtaining a minimum measurement sensitivity of $\delta\Omega_{\min,a,squeezed} = 3.8 \times 10^{-5} Hz$. The investigated structure will enable compact miniaturization and high sensitivity inertial rotation sensors to become more achievable in practice.

### Funding

Financial support by National Natural Science Foundation of China (Grant Nos. 61975005, 51872010 and 11804017) and Beijing Academy of Quantum Information Sciences (Grant Nos. Y18G28)

### Acknowledgments

We thank Prof. Guofeng Zhang for the useful comments and suggestions.